# Achieving Giant Magneto-optic Effects with Optical Tamm States in Graphene-based Photonics

Haixia Da[1], Cheng-Wei Qiu[1], Qiaoliang Bao[2], Jinghua Teng[3], Kian Ping Loh[4,5], and Francisco J. Garcia-Vidal[6]

[1]*Electrical and Computer Engineering Department, National University of Singapore, 4 Engineering Drive 3, Singapore 117576*

[2] *Department of Materials Engineering, Monash University, Clayton Campus, Clayton, Victoria 3800, Australia*

[3]*Institute of Materials Research and Engineering, Agency for Science, Technology, and Research (A\*STAR), 3 Research Link, Singapore 117602*

[4] *Graphene Research Centre, National University of Singapore, 2 Science Drive 3, Singapore 117542, Singapore.*

[5]*Department of Chemistry, National University of Singapore, 3 Science Drive 3, Singapore 117543*

[6]*Departamento de Fisica Teorica de la Materia Condensada, Universidad Autonoma de Madrid, 28049 Madrid, Spain.*

ABSTRACT

We manipulate optical Tamm states in graphene-based photonics to achieve and steer large magneto-optical effects. Here we report the presence of a giant Faraday rotation via a single graphene layer of atomic thickness while keeping a high transmission. The Faraday rotation is enhanced across the interface between two photonic crystals due to the presence of an interface mode, which presents a strong electromagnetic field confinement at the location of the graphene sheet. Our proposed scheme opens a promising avenue to realize high performance graphene magneto-optical devices that can be extended to other two-dimensional structures.

Graphene has attracted considerable interest due to its unique electronic properties such as anomalous quantum Hall effect, Klein paradox and electronic lens in p-n junction, as well as its potential applications in this area as field-effect transistors or quantum reflective switches[1-4]. In addition, graphene has also been demonstrated to be a good candidate in optoelectronics[5] with potential applications in photodetection, photovoltaics, transformation optics and plasmonics[6-14]. Another promising application of graphene is in magneto-optics. The opportunity of creating giant Faraday rotation (FR) angle has been identified experimentally in both single layer and multilayered graphene structures, which provides completely new candidates for producing magneto-optical (MO) effects[15-19]. The electromagnetic non-reciprocity and gyrotropy of a single layer graphene under the application of a magnetic field have also been demonstrated [20]. FR angle in a two-dimensional system such as graphene with infinitesimal thickness is a fascinating physical phenomenon which distinguishes from that observed in conventional bulky magnetic materials, in which the FR angle is proportional to the distance that light travels. The reported experimental FR angle for a free standing single pristine or patterned graphene layer itself does not exceed 0.15 rad, which does not satisfy the demand of MO devices. On the other hand, graphene sheets embedded in cavities have been found to be very effective in achieving a larger FR angle[16-17]. However, due to the tradeoff between FR angle and transmission, transmission rates of those reported graphene devices are usually low, which restricts their practical applications. It is then vital to find alternative routes to improve MO effects in graphene while keeping a high transmission of light through the structure.

Tamm states refer to interface electron states occurring within an energy band gap and spatially located at a crystal surface[21]. Since the original concept proposed by Tamm and co-workers in 1932, there have been extensive studies on these modes in conventional semiconductors or metals[22-24]. In analogy with the electronic case, optical Tamm states (OTMs) have been explored in different dielectric photonic crystals and metamaterials[25-28]. OTMs have also greatly contributed to the development of the field of magneto-optics as shown, for example, in the bulky magnetic materials analyzed in Ref. 29. It is shown that the improved MO performance in these structures is due to the existence of OTMs located at the interface between the two sub-systems. In contrast, such an effect is absent in simple periodic or aperiodic magneto-photonic crystals due to the above-mentioned intrinsic tradeoff between the transmission (reflectance) and FR (Kerr) effects[30-31].

In this Letter, we report simultaneously improved FR angle and transmission in a graphene-based heterogeneous photonic crystal (G-HPC). By appropriately manipulating the parameters of the two types of photonic crystals (PCs), the amplitude of the FR angle is estimated to be as large as $-10.9°$ at the specific operating frequency of 25 THz. This enhancement is about two orders larger than that reported in HPC structures with bulk magnetic materials[29]. The mechanism behind the good MO performance is revealed by the combination of an effective refraction model and the analysis of the electromagnetic field distribution. Additionally, the spectral position of the large FR angle in graphene, which is intrinsically sensitive to the external magnetic field via a square-root dependence, is suppressed in G-HPCs due to the resonant character of the OTMs involved.

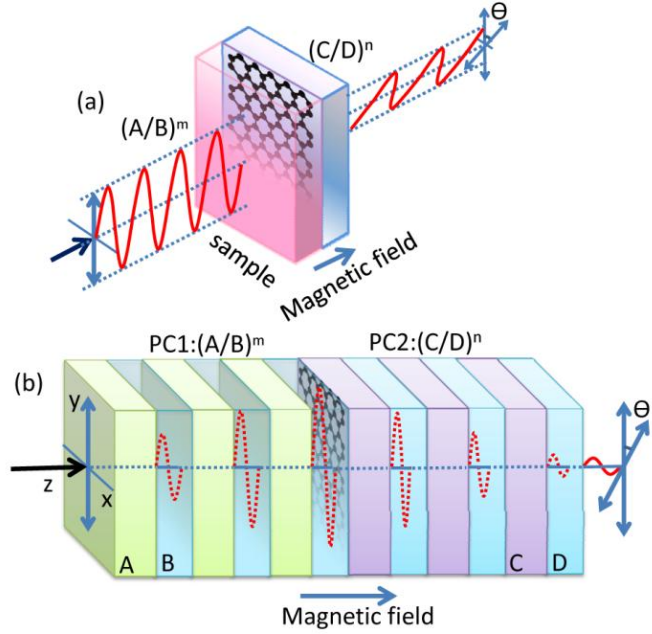

**Figure 1.** (a). Schematic diagram of the experimental setup used to measure FR angle, where the sample is a graphene-based structure and an external magnetic field is applied in the direction perpendicular to the graphene sheet. (b). The proposed HPC structure with graphene, i.e., $(A/B)^m$/graphene/$(C/D)^n$, where A, B, C and D represent isotropic dielectric layer and m (n) is their repetition number in PC1 (PC2), respectively.

A schematic picture of the proposed structure (G-HPC) is shown in Fig. 1a. The single-layer graphene lies in the *x-y* plane and an external magnetic field (B) is applied along the *z* direction. A linearly *p*-polarized monochromatic electromagnetic wave impinges normally onto the configuration. Fig.1b shows the design of a single layer graphene embedded into the HPC, where the graphene is sandwiched between the two PCs. Label A, B, C and D represent four isotropic dielectric materials and m and n are the repetition numbers of the A/B and C/D subsystems, respectively. For proof of principle purposes, in our calculations we consider silicon as the dielectric material for both A and C components whereas SiC is the one used for both B and D.

The refractive indexes of Si and SiC for the range of frequencies analyzed are 1.5 and 2.55, respectively[32]. The thicknesses of the four PC components are taken as $d_A$=4.99 µm, $d_B$=1.39 µm; $d_C$=0.95 µm and $d_D$=3.77 µm in order to operate at the specific working frequency of 25 THz. The chosen thicknesses of the components are related to the existence of an OTM acquired by the phase matching condition $r_L r_R = 1$ [33], where $r_L$ and $r_R$ are reflection coefficients of the left and right periodic structures with wave impinging from vacuum, respectively (see Supplemental Material for details). Regarding the optical properties of graphene, we operate in the so-called quantum regime by taking $e_F = 4$ meV. We assume that the scattering rate is $\Gamma$=1.2 meV and the temperature is 10 K when introducing its AC-conductivity into the calculations[34].

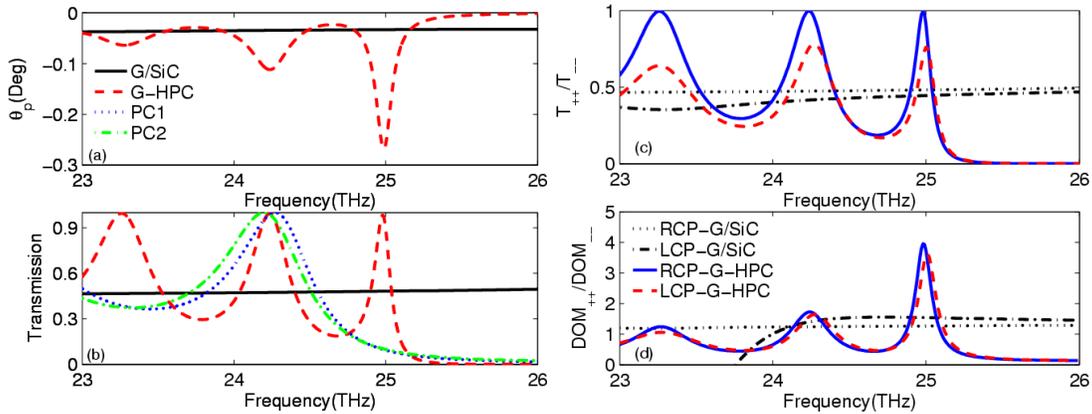

**Figure 2.** (a) FR angle as a function of frequency in the two-layer structure composed of graphene and SiC (G/SiC), HPC structure with graphene (G-HPC) and two bare PCs, where m=n=6. (b) Corresponding transmission coefficient for the four structures. In the G-HPC configuration, a FR dip and a sharp transmission peak appearing within the band gap of the two PCs are observed. (c) The transmission and (d) the corresponding density of modes (DOM) in arbitrary units for RCP and LCP lights of conventional G/SiC and the proposed G-HPC structure, respectively.

Calculations of the FR angle and transmission were performed by employing the transfer matrix method as described in Ref. 16. Figs. 2a and b show the FR angle and transmission at a given magnetic field B=1 T in the proposed G-HPC structure. The results of the two-layer structure (graphene monolayer on top of a SiC substrate of thickness 1.39 μm), PC1 and PC2 are also shown for comparison. It is found that there is a pronounced enhancement of the FR angle, $\theta_p$, at the specific frequency of 25 THz for the G-HPC structure. Its spectral location falls within the frequency band gap region of two PCs. Apparently, FR angle is zero for the PC1 and PC2 structures due to reciprocity of all their components. These results indicate that the large MO effects in the G-HPC structure originates from the combined effect of the graphene layer and cavity formed by the two PCs. We can see that the peak value of FR angle is -0.27° and the corresponding transmission is very high, 0.98. The FR angle shown here is about 8 times larger than that of the two-layer structure and it also yields a greatly increased transmission, going from 0.48 to 0.98. Thus the important feature here is that the enhancement of the FR angle and a higher transmission are achieved simultaneously, indicating the MO performance of the graphene sheet is significantly improved.

It is instructive to inspect the quantitative mechanism that dominates the enhancement of FR profiles in the G-HPC structure. This can be understood in terms of the effective circular birefringence of the hybrid structure. To this end, we use the retrieval method [35] to construct the relationship between the optical conductivity of graphene and the effective parameters of the two PCs from the effective medium point of view. Details of this derivation can be found in the Supplemental Material. The important asset of this approach is that we can work with an analytical formula (Eq. S3) for the ratio between the right circularly polarized (RCP) light and left circularly polarized (LCP) light transmission coefficients, $t_{++}/t_{--}$. Using the above-mentioned formula, the transmission spectra for both RCP ($T_{++} = |t_{++}|^2$) and LCP ($T_{--} = |t_{--}|^2$) lights can be calculated and are shown in Fig. 2c. The result of the G/SiC structure is also shown, which has no prominent features, while there are three resonances in the G-HPC transmission spectra over the frequency region we have studied for both RCP and LCP lights. At the three resonances, the peak of RCP transmission amplitude approaches almost 1 and is larger than that of LCP light. Moreover, the RCP and LCP peaks acquire the maximum values at slightly different frequencies, indicating a different phase modulation of the G-HPC structure for the two different polarizations. We can correlate the three resonant features in the $T_{++}/T_{--}$ spectrum with the density of optical modes (DOM) in the structure. This physical parameter carries information of the availability of allowed photonic states within a certain frequency range from environmental influence, which enters into the electromagnetic dynamics via the derivative of photonic dispersion relationship[36]. In our model, DOM is determined by the derivative of the effective wave vector parallel to the propagating direction in our finite system[37]. Fig. 2d shows the DOM versus frequency for RCP and LCP components. The DOM value of the G/SiC structure is enlarged by 100 times for visualization at the same scale. As expected, the G/SiC structure does not exhibit any resonant effect, thus leading to broadband small FR angle. In the G-HPC structure, the DOM of RCP and LCP lights are observed to oscillate over the 23-26 THz frequency range. Two small DOM peaks appear at frequencies corresponding to a Fabry-Perot resonance (23.3 THz) and band-edge transmission resonance (24.3 THz). DOM achieves the maximum value at the specified working frequency 25 THz, which is related to the significant decrease of the group velocities and thus the strongly spatial localization of light at the interface between the two PCs. Besides, slightly different localization conditions emerge around the specified working frequency for effectively RCP and LCP lights due to their different propagation velocities. Consequently, the designed heterogeneous PC structure with the interface mode has an intrinsic contribution to modulate the effective RCP and LCP lights in the G-HPC structure, thus generating a large FR angle.

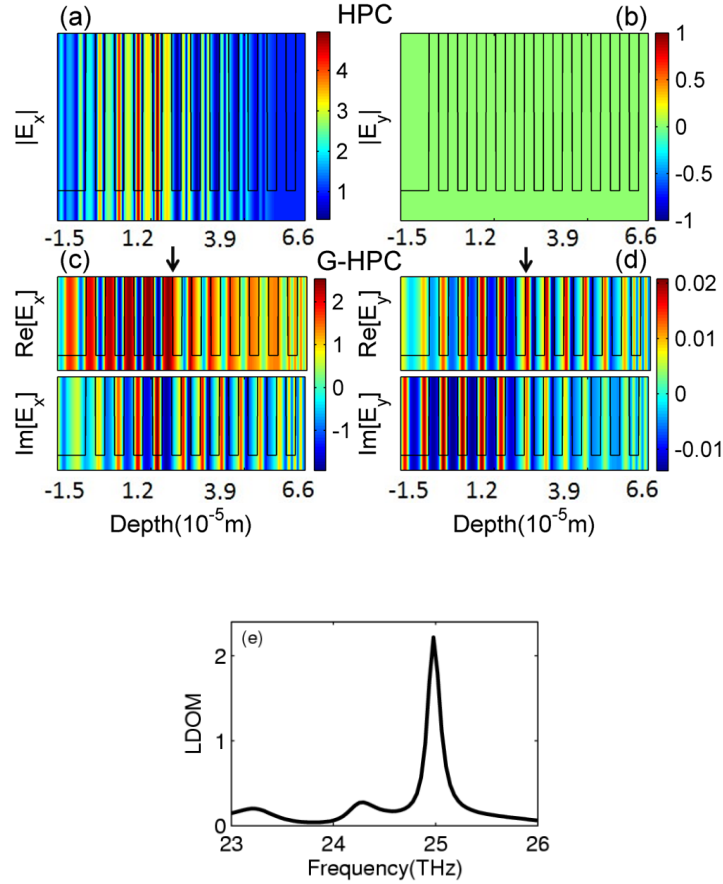

**Figure 3.** The electric field intensity distribution $|E_x|$ (a) and $|E_y|$ (b) across the HPC structure without graphene; The real and imaginary parts of electric intensity distribution $E_x$ (c) and $E_y$ (d) across the HPC structure with graphene. The arrows show the location of graphene. The intensities of the electric and magnetic fields evaluated at the interface between the two PCs are strong, where the resonant tunneling appears. The existence of $E_y$ component indicates the rotation in the polarization of transmitted wave when graphene is sandwiched between the two PCs. The solid line represents the corresponding refractive index profile. (e) The local density of mode (LDOM) in arbitrary units as a function of frequency calculated at the specific location where the graphene sheet is placed.

It is now interesting to visualize the role of the HPC structure in the enhancement of graphene's MO performance. We have calculated the local field intensity distribution across the whole structure at the excitation frequency of 25 THz. In Fig. 3, we plot the spatial profiles of the electric field of the HPC structure with and without graphene. Panels (a) and (b) present the electric field distributions of $|E_x|$ (a) and $|E_y|$ (b) for the HPC configuration without a graphene layer in between. The most important feature is that there is no $E_y$ component in the absence of graphene, indicating that no propagating wave gets rotated in the polarization plane of the incident wave. Furthermore, one can observe that the electric (and magnetic as well, not shown) fields reach local maxima at the center of each A/B component within PC1 rather than at the interfaces of the A-B bilayers. There are two local maxima also within each component of the PC2 and the $E_x$ component of the field intensity of the wave in PC2 is overall smaller than that in PC1. Close to the interface between PC1 and PC2, the electric field reaches the global maximum value. Both the electric and magnetic fields are exponentially decaying within the two PCs away from the interface, which indicates that a standing wave emerges allowing the resonant tunneling of light through the whole structure.

In contrast to the HPC structure without graphene, Fig. 3c and 3d show that the propagating wave through the G-HPC structure has not only an $E_x$ component but also an $E_y$ component, despite the fact that the incident wave had only an $E_x$ component. The electric field enhancement in $E_x$ and $E_y$ components appear in an alternating pattern within the two PCs. It reveals that the polarization plane of the transmitted wave gets rotated due to the gyrotropy effect of graphene under the application of B field. Besides, one can see that the electric field amplitude reaches its maximum near to the interface between the two PCs and the decaying wave penetrates into several layers into the two PCs. The field intensity is localized, to a large extent, at the center of the whole structure where the single atomic layer graphene is located, suggesting that the enhanced MO effects originate from the electromagnetic field localization. The strong interference between forward and backward propagating waves leads again to a standing wave and thus a robust localized interface mode, the OTM, which increases the coupling between the incident wave and graphene. This resonant tunneling of an electromagnetic wave is empowered by OTMs that determines the enhanced FR angle and its associated high transmission. The FR angle can be roughly estimated from the ratio of $E_y$ and $E_x$ components of the transmitted light. Therefore, the FR angle per unit of wave propagation length along the G-HPC structure is obviously large even for a single atomic graphene layer due to the confinement of resonant states. The key role played by the OTM is illustrated in Fig. 3e, which shows the local density of modes (LDOM) versus frequency evaluated at the location of the graphene sheet. A close correspondence between the spectral locations in which FR is enhanced and the resonant peaks in the LDOM is observed.

In the quantum regime, the Dirac character of the electrons in graphene leads to a cyclotron resonance that increases with the B field following a square-root dependence[34]. Therefore, it provides us with another degree of freedom to further enhance the MO effects together with high transmission, using our new scheme. We investigate the influence of this factor on MO effects in the G/SiC and the G-HPC structures. Figs. 4a and 4c represent the FR angle, $\theta_p$, and transmission of the G/SiC structure when the magnetic field varies from 1 to 7 T. It can be seen that there is a dip in FR angle for each individual B field and that the dip's position is shifted to higher frequencies following the square-root dependence of the cyclotron frequency with the B field. The corresponding transmission spectra evaluated at the different B fields are not sensitive to the B field and transmission is always less than 0.65 for the maximum FR angle. Figs. 4b and 4d show the FR angle, $\theta_p$, and transmission of the proposed G-HPC structure, respectively. Our results imply that the increase of the B field further enhances $\theta_p$ (-10.9° at B=7 T) but does not degrade the transmission too much, i.e., it decreases from 0.98 to 0.85 when the B field increases from 1 to 7 T.

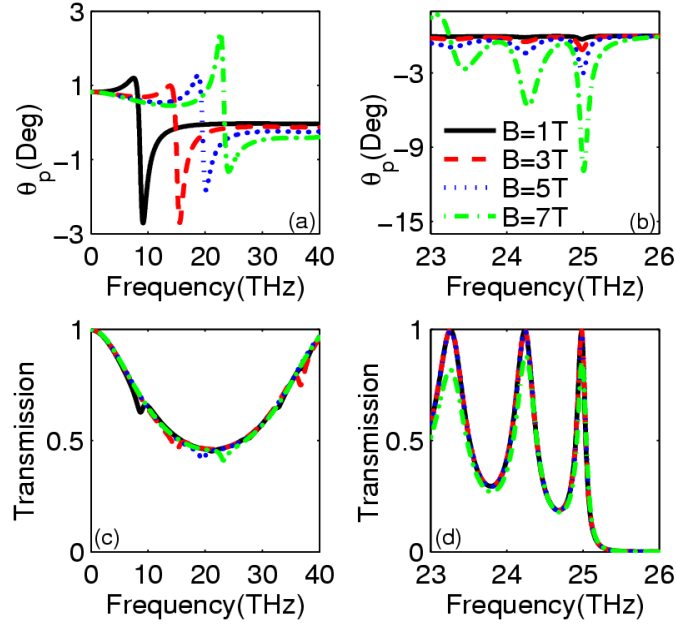

**Figure 4.** (a) FR angle and (c) transmission are plotted as a function of frequency for the G/SiC structure; (c) FR angle and (d) transmission as a function of frequency for the G-HPC structure. There is a dip in the FR angle, whose position varies with B field. The enhanced FR angle is numerically predicted with B field after the HPC structure is employed but the dependence of the position of FR dip on B field is suppressed due to the fixed geometrical structure.

The enhancement of $\theta_p$ with the B field can be traced back to the relativistic Landau levels quantization of graphene[1]. The value of FR angle is closely related to the real part of the off-diagonal term of optical conductivity $\sigma_{xy}$, Re[$\sigma_{xy}$]. There is only one peak appearing in the spectrum of Re[$\sigma_{xy}$] within the frequency region we investigate. Importantly, the peak value increases with the magnitude of B field. The combination of this fact with the confined character of the interface OTM results in a further enhancement in $\theta_p$ with the B field. The position of the enhanced $\theta_p$ is found to be insensitive to the magnitude of the B field. This is markedly different from the case of the two-layer structure, in which the position of FR dip varies with B field. This distinction may be understood because the enhanced FR angle in the G-HPC structure is mainly determined by spectral location of the OTM that has a structural/dielectric origin and not by the cyclotron resonance.

According to our results and modeling, the G-HPC structure can enhance the MO performance of graphene at a desirable working frequency given the proper geometric arrangement of the two PCs. It is worth noticing that we have used just two different dielectrics and a very limited number of periods. The mechanism could be applied to other frequency ranges by simply modifying the structural parameters. For example, if we want to operate at lower frequencies (for example at 10 THz), the improved MO effect can be achieved using the same two types of dielectrics (Si and SiC) as in previous calculations but changing the structural parameters to $d_A$=4.20μm, $d_B$=2.61μm; $d_C$=3.62μm and $d_D$=2.95μm. For this operating frequency (10 THz), FR angle increases from 0.28° further to 5.58° and the transmission increases from 0.476 to 0.9763 at B=1T when going from a two-layer structure to the G-HPC system. This unique OTM-assisted MO effect enables us to find wide applications for other ultra-thin magnetic materials, since one can have both large FR angle and high transmission. Finally, we have also investigated the performance under oblique incidence and it is found that the simultaneous enhancement in FR angle and high transmission is still present while slightly smaller than in the case of normal incidence.

In summary, we have examined the FR angle in a heterogeneous photonic crystal with a single layer graphene placed at the interface, where a giant rotation angle and high transmission are achieved in the quantum regime of graphene. Such counter-intuitive improvement is supported by the existence of OTM occurring at the interface between the two photonic crystals. The FR angle is estimated to be as large as -10.9° at the chosen operating frequency. This value is about two orders larger than that in conventional photonic crystals made of bulky magnetic materials. Importantly, the magnitude of FR peak increases with the external magnetic field but its spectral location is determined by the OTM and thus insensitive to the magnetic field. These features of simultaneously enhanced MO effects and high transmission are markedly distinguished from the currently existing graphene devices.

**Supplemental Material**

Details of the condition of OTM existence, effective refraction model to derive magneto-optical effects, transmission, and the effect of the stacking sequence to MO effects in graphene-based heterogeneous photonic crystals are discussed.

**ACKNOWLEDGMENT**

This work was supported by the TDSI/11-004/1A and IMRE/12-1P0903. We thank Dr. Ding Weiqiang for useful discussions. F.J.G.-V. acknowledges financial support from the Spanish MINECO under contract MAT2011-28581-C02-01.